\documentclass[a4paper,11pt]{article}
\usepackage{graphicx}
\usepackage{graphics}
\def\G{{\tilde G}}

\title{\bf  Observables in  elastic
electron-deuteron scattering with two-photon exchange}

\author{\bf Yu Bing Dong$^{1,2}$,
Chung Wen Kao$^3$, Shin Nan Yang$^4$, and Yu Chun Chen$^4$\\
$^1$CCAST (World Laboratory), 100080 Beijing, P. R. China\\
$^2$Institute of High Energy Physics, Chinese Academy of Sciences,
Beijing, P. R. China\\
$^3$Department of Physics, Chung-Yuan Christian University, Chung-Li
320, Taiwan\\$^4$ Department of Physics and Center for Theoretical
Sciences,
National Taiwan University,\\
Taipei 10617, Taiwan\\
}
\begin{document}
%\date{}
\maketitle
\begin{abstract}
The general form of the cross section as well as the polarizations
of electron-deuteron elastic scattering are given. Two-photon
exchange effects are analyzed. Possible signatures of the two-photon
exchange effects in the
electron-deuteron elastic scattering are discussed. \\
\end{abstract}
 \par {\hskip 0.1cm}Two-Photon exchange.

\newpage
\section{Introduction}

\par\noindent\par\vspace{0.2cm}

Electromagnetic form factors of the nucleon provide us with
essential information about the internal structure of the nucleon as
they describe the distribution of charge and magnetization inside
the nucleons. Traditionally, the Sachs electric $G_E(Q^2)$ and
magnetic $G_M(Q^2)$ form factors are determined from the $eN$
elastic scattering by the Rosenbluth separation method. Namely, in
the one-photon exchange (OPE) approximation, the unpolarized $eN$
cross section is a linear function of photon polarization parameter
$\epsilon$ for a given value of four-momentum transfer squared
$Q^2$,
\begin{eqnarray}
d\sigma_0=A_0\Big ( G_M^2(Q^2)+\frac{\epsilon}{\tau_N}
G_E^2(Q^2)\Big ), \label{Rosenbluth}
\end{eqnarray}
where the factor $A_0$ depends on kinematic variables, and
$\tau_N=Q^2/4M^2_N$ with $M_N$ the nucleon mass. $G_M(Q^2)$ and
$G_E(Q^2)$ can then be determined from the intercept and the slope
from the linear plot of $d\sigma_0 \,\,vs.\,\, \epsilon$. The ratio
$R=\mu_pG_E/G_M$ determined by the Rosenbluth separation method from
the $ep$ scattering, where $\mu_p=2.79$ is the proton magnetic
moment, has been consistent with $R \approx 1 $ for $Q^2
< 6$ GeV$^2$ \cite{Arrington03}.\\

Another way to measure $R$ is with the use of polarized transfer
technique. In the Born approximation for the elastic $ep$
scattering, a longitudinally polarized electron transfers its
polarization to the recoil proton with two nonzero components,
$P_t$, perpendicular to, and $P_l$ parallel to, the proton momentum
in the scattering plane. Simultaneous measurements of these two
polarizations give \cite{Akhiezer74}
\begin{eqnarray}
\frac{P_t}{P_l}=-\sqrt{\frac{2\epsilon}{\tau_N(1+\epsilon)}}\frac{G_E}{G_M}.
\label{Rpol}
\end{eqnarray}
Recent polarization transfer experiments at Jefferson Lab
\cite{Jones00,Gayou02} give, however, a result $R\simeq
1-0.135(Q^2-0.24)$, which differs substantially from $R \approx 1 $
as obtained by the Rosenbluth separation method  over the same range
of $Q^2$
\cite{Arrington03}, exhibiting a nonscaling behavior.\\

Two-photon exchange effects have been proposed to account for the
discrepancies between the electromagnetic form factors measured
through the Rosenbluth separation method and the polarization
transfer method \cite{Blunden03,Guichon03,Chen04}. In Ref.
\cite{Blunden05}, the two-photon exchange contributions to the
neutron electromagnetic form factors have been estimated and found
to be important at large $Q^2$. It is hence natural to expect that
the two-photon exchange mechanisms could also give a non-negligible
contribution to the electron-deuteron elastic and quasi-elastic
scattering. The question of the two-photon exchange (TPE) effects in
the electron-deuteron ($eD$) elastic scattering has been studied in
Refs. \cite{Gunion73,Franco73,Boitsov73,Lev75,Rekalo99}, focusing on
the possible signatures in the elastic cross sections. It is
possible that the TPE effects in the $eD$ scattering might be more
easily accessible in the polarization observables, just as in the
case of $ep$ scattering~\cite{Wells01,Maas05}. It is hence important
to go beyond the Born approximation and derive the expressions for
all observables, including the polarizations, in the elastic
electron-deuteron scattering to provide a theoretical framework for
analyzing data and the extraction of the deuteron form factors
beyond the OPE approximation in a model-independent way. These
general expressions can also be useful in choosing certain
kinematical regions and combinations of observables which would be
more sensitive to the TPE effects.
\\

In this article, we present a general formulation of the elastic
electron-deuteron scattering with multiple photon exchanges and
derive the formulas for all possible observables, including
unpolarized differential cross section, three vector polarization
observables $P_x, P_y$ and $P_z$ and three tensor polarization
observables $T_{20}$, $T_{21}$ and $T_{22}$.  These formulas are
given in Sec. 2. In Sec. 3, we examine the expressions obtained in
Sec. 2 and explore the possibilities where
the TPE effects might most likely be observed and summarize in Sec. 4.\\

\section{Electron-deuteron elastic scattering with multi-photon exchange}
\par\noindent\par

The electromagnetic form factors of the deuteron are defined by the
matrix elements of the electromagnetic current $J_\mu(x)$ according
to
\begin{eqnarray}
<p_4,~\lambda'\mid J_\mu(0)\mid p_2,~\lambda> &=&-e_D\bigg\{\Big
[G_1(Q^2) \xi'^*(\lambda')\cdot \xi(\lambda)
-G_3(Q^2)\frac{(\xi'^*(\lambda')\cdot q)(\xi(\lambda)\cdot
q)}{2M^2_D}\Big ]\cdot
P_\mu \nonumber \\
&&+G_2(Q^2)\Big [\xi_{\mu}(\lambda)(\xi'^*(\lambda ')\cdot q)-
       \xi'^*_{\mu}(\lambda ')(\xi(\lambda)\cdot q)\Big ]\bigg \},
       \label{current}
\end{eqnarray}
where $p_4,\xi',\lambda'$ and $p_2,\xi,\lambda $ denote the
momentum, helicity, and polarization vector of the final and initial
deuterons, respectively. $q=p_4-p_2$ is the photon momentum,
$P=p_2+p_4,\, Q^2=-q^2$ the four-momentum transfer squared, $M_D$
the deuteron mass and $e_D$ is the charge of the deuteron. In the
one-photon exchange approximation or Born approximation, the
unpolarized differential cross section of the elastic
electron-deuteron scattering
\begin{eqnarray}
e(p_1,s_1)+D(p_2,\xi)\rightarrow e(p_3,s_3)+D(p_4,\xi')
\label{eDscatt}
\end{eqnarray}
in the laboratory frame is given by~\cite{Jankus56}
\begin{eqnarray}
\frac{d\sigma}{d\Omega}=\frac{d\sigma}{d\Omega}\bigg
|_{Mott}I_0(OPE), \label{diffCrx}
\end{eqnarray}
with
\begin{eqnarray}
I_0(OPE)=A(Q^2)+B(Q^2)tan^2\frac{\theta}{2}, \label{I0-ope}
\end{eqnarray}
where $\theta$ is the scattering angle of the electron,
$(d\sigma/d\Omega)_{Mott}$ is the Mott cross section for a
structure-less particle
\begin{eqnarray}
\frac{d\sigma}{d\Omega}\bigg |_{Mott}=\big( \frac{\alpha}{2E}\big
)^2 \frac{\cos^2\frac{\theta}{2}}{\sin^4\frac{\theta}{2}}
\frac{1}{1+\frac{2E}{M_D}\sin^2\frac{\theta}{2}}
=\sigma_0\cot^2\frac{\theta}{2}, \label{sigma-Mott}
\end{eqnarray}
and
\begin{eqnarray}
A(Q^2)=G_c^2(Q^2)+\frac23\tau
G_M^2(Q^2)+\frac89\tau^2G_Q^2(Q^2),~~~~~~~
B(Q^2)=\frac43\tau(1+\tau)G^2_M(Q^2). \label{AB}
\end{eqnarray}
In Eqs. (\ref{diffCrx}-\ref{AB}), $\alpha$ is the fine structure
constant, $\tau=Q^2/4M^2_D$, and $E$ is the incident electron
energy. $G_M$, $G_C$ and $G_Q$ are, respectively, the deuteron
magnetic, charge and quadrupole form factors. The relations between
$G_M$, $G_C$ and $G_Q$ and the form factors $G_1$, $G_2$ and $G_3$
defined in Eq. (3) are
\begin{eqnarray}
G_M=G_2, ~~~~G_Q=G_1-G_2+(1+\tau )G_3,~~~ G_C=G_1+\frac23\tau G_Q.
 \label{GMCQ_G123}
\end{eqnarray}
The normalizations of the three form factors are, respectively,
$G_c(0)=1,~G_Q(0)=M^2_DQ_D=25.83$, and $G_M(0)=1.714$.\\

Note that in the well-known Rosenbluth separation of Eq. (6), there
are two unpolarized structure functions $A$ and $B$, and three
independent form factors $G_C$, $G_Q$ and $G_M$, for the deuteron, a
spin-one particle. To determine the three form factors completely,
one thus needs at least, one polarization observable. The optimal
choice, in the literature, is the polarization $T_{20}$ (or
$P_{zz}$). This is because three tensor polarizations, like
$T_{20}$, $T_{21}$ and $T_{22}$, can be measured in the $eD$
scattering with unpolarized beam and target while the  observables
like $T_{21}$ and $T_{22}$ are small in magnitude because they are
proportional to $G_M$.\\

It is known that Lorentz, parity, charge-conjugation invariance
dictate that the T-matrix for the elastic electron-nucleon can be
expanded in terms of six independent Lorentz structures. In the
limit of zero electron mass, helicity conservation would reduce them
to three~\cite{Guichon03}. The remaining three structure functions
are all complex. In the OPE approximation, these three complex form
factors further simplify to two real form factors $G_M(Q^2)$  and
$G_E(Q^2)$. In the case of electron-deuteron elastic scattering,
there are 36 Lorentz invariant amplitudes which reduce to 18 if
parity is conserved. With charge-conjugation invariance, only 9 of
them are independent. In the limit of zero electron mass, i.e.,
helicity is conserved, then one ends up with only 6 independent
amplitudes. In analogy to virtual Compton scattering (virtual photon
$\rightarrow$ deuteron, proton $\rightarrow$ electron), we can
express the most general form of the $eD$ elastic scattering as
\cite{Tarrach75}
%including the OPE (${\cal C}=-1$) and TPE (${\cal C}=+1$) takes the form
\begin{eqnarray}
{\cal M}^{eD}=-e^2\bar{u}(p_3,s_3)
\gamma_{\mu}u(p_1,s_1)\frac{1}{q^2}\sum_{i=1}^6\G_iM_i^{\mu},
\label{MeD}
\end{eqnarray}
where
\begin{eqnarray}
M_1^{\mu}&=&(\xi'^*\cdot\xi)P^{\mu},\nonumber \\
M_2^{\mu}&=&\Big [\xi^{\mu}(\xi'^*\cdot q)
-(\xi\cdot q)\xi'^{*\mu}\Big ],\nonumber \\
M_3^{\mu}&=&-\frac{1}{2M_D^2}(\xi\cdot q)(\xi'^*\cdot q)P^{\mu},\nonumber \\
M_4^{\mu}&=&\frac{1}{2M_D^2}(\xi\cdot K)(\xi'^*\cdot K)P^{\mu},\nonumber \\
M_5^{\mu}&=&\Big [\xi^{\mu} (\xi'^*\cdot K)
+(\xi \cdot K)\xi'^{*\mu}\Big ],\nonumber \\
M_6^{\mu}&=&\frac{1}{2M_D^2}\Big [(\xi\cdot q)(\xi'^*\cdot K)
-(\xi\cdot K)(\xi'^{*}\cdot q)\Big ]P^{\mu}. \label{M1-6}
\end{eqnarray}
with $K=p_1+p_3$. Generally speaking, the form factors $\G_i$ with
$i=1,6$, are complex functions of $s=(p_1+p_2)^2$ and
$Q^2=-(p_1-p_3)^2$. They can be expressed as
\begin{eqnarray}
\G_i(s,Q^2)=G_i(Q^2)+G_i^{(2)}(s,Q^2), \label{Gtilde}
\end{eqnarray}
where $G_i$'s correspond to the contributions arising from the
one-photon exchange and $G_i^{(2)}$'s  stand for the rest which
would come mostly from the TPE. In the OPE approximation,
$G_4=G_5=G_6=0.$ It is easy to see that $G_i$ $(i=1,2,3)$ is of
order of $(\alpha)^0$ and $G_i^{(2)}$ ($i=1,...6$) are
of order  $ \alpha $. \\

It is now straightforward though tedious to derive the unpolarized
differential cross section for the $eD$ elastic scattering with the
general form of the scattering amplitude of Eqs.
(\ref{MeD}-\ref{M1-6})  which includes the TPE effects. If the
contributions from pure TPE are neglected and only the interference
terms between the OPE and TPE contribution are retained, the
resulting differential cross section can be expressed as
\begin{eqnarray}
\frac{d\sigma}{d\Omega}&=&\frac{d\sigma}{d\Omega}\bigg |_{Mott}I_0\nonumber \\
&=&\frac{d\sigma}{d\Omega}\bigg |_{Mott} \bigg  \{\Big [(A+\Delta
A)+(B+\Delta B) \tan^2\frac{\theta}{2}\Big]
+\Delta\sigma(\theta, Q^2)\bigg \}\nonumber \\
&=&\sigma_0 \bigg \{\Big [(A+\Delta
A)\cot^2\frac{\theta}{2}+(B+\Delta B)\Big ] +\Delta
\sigma(\theta,Q^2) \cot^2\frac{\theta}{2}\bigg \}, \label{dsigma-eD}
\end{eqnarray}
where $A$ and $B$ are the same as the ones in Eq. (\ref{AB}). In Eq.
(\ref{dsigma-eD}), $\Delta$ is used to indicate the contributions
coming from the interference terms between the OPE and TPE which
have not been considered before
\cite{Arnold80,Garcon90,Alexa99,Garcon94}. $\Delta A$ and $\Delta B$
are given as
\begin{eqnarray}
\Delta A&=& 2\Big [G_cRe(G_C^{(2)*})+\frac23\tau G_MRe(G_M^{(2)*})
+\frac89\tau^2G_QRe(G_Q^{(2)*})\Big ]\nonumber \\
&&+\frac{4\tau^2}{3}\Big [(2\tau+1)G_1
-2(\tau+1)G_2+2\tau(\tau+1)G_3\Big ]Re(G_4^{(2)*}),\nonumber\\
\Delta B&=&\frac83\tau(1+\tau)G_MRe(G_M^{(2)*}), \label{DeltaAB}
\end{eqnarray}
and
\begin{eqnarray} \Delta \sigma(\theta,Q^2)&=&\frac{2}{3} \bigg
\{2\tau \cot^2\frac{\theta}{2}\Big [ (2\tau-1)G_1-2\tau G_2
+2\tau^2G_3\Big ] Re(G_4^{(2)*})\nonumber \\
&&+\frac{K_0}{M_D}\Big [\Big ( (2\tau-1)G_1-2\tau G_2
+2\tau^2G_3-2\tau \tan^2\frac{\theta}{2}G_2\Big )Re(G_5^{(2)*})\nonumber \\
&&+2\tau \Big ( (2\tau+1)G_1-(2\tau+1)G_2+2\tau(\tau+1)G_3\Big )
Re(G_6^{(2)*})\Big  ] \bigg \}, \label{Delta-sigma}
\end{eqnarray}
where
\begin{eqnarray}
K_0^2=(p_{10}+p_{30})^2=4M^2_D\tau\left [(1+\tau)
+\cot^2\frac{\theta}{2}\right ]. \label{K0}
\end{eqnarray}
Note that $\Delta A$ and $\Delta B$ contain no explicit
$\theta$-dependence, if all of $G^{(2)}_{i},i=1-6$ would be
independent of $\theta$, while the last term in Eq.(\ref{dsigma-eD})
contains explicit $\theta$-dependence through $\Delta\sigma$ besides
the factor $\cot^2(\theta/2)$. However, according to Eq.
(\ref{Gtilde}), $G^{(2)}_{i}$ are functions of $\theta$ and $Q^2$,
because $s=s(\theta,Q^2)$ in the laboratory frame. Therefore $\Delta
A$ and $\Delta B$ both depend on the $\theta$ through $G^{(2)}_{i}$.
  \\

We next turn to the  polarization observables which enter in the
$eD$ elastic scattering. The relations between the notation of
Arnold {\it et al.} \cite{Arnold80} and the popular one of Garcon
and Orden~\cite{Garcon90} are
\begin{eqnarray}
&P_{zz}=\sqrt{2}Re(T_{20}),  &P_{xz} =-\sqrt{3}Re(T_{21}),
\hspace{0.8cm}
 (P_{xx}-P_{yy}) = 2\sqrt{3}Re(T_{22}),\nonumber \\
&P_{z} = -\sqrt{\frac 23}Re(T_{10}),
&P_y=-\frac{2\sqrt{3}}{3}Im(T_{11}), \hspace{0.8cm}P_x
=-\frac{2\sqrt{3}}{3}Re(T_{11}). \label{pol}\end{eqnarray} The
tensor polarizations in the $eD$ unpolarized elastic scattering are
$T_{20}$ ($P_{zz}$), $T_{21}$($P_{xz}$), and $T_{22}$
($P_{xx}-P_{yy}$) (the convention of Ref. \cite{Arnold80}). In this
paper, we use the notation of Ref. \cite{Arnold80}. There are
several ways to extract those polarization observables. For example,
the vector polarizations  $P_x$ and $P_z$ can be measured by using
the longitudinal polarized electron beam and the unpolarized
deuteron target \cite{Arnold80}. They are given by the the asymmetry
of the cross section with the different polarizations of the
electron beam. The vector polarization $P_y$ results from the vector
polarized final deuteron along $y$ direction which is perpendicular
to the scattering plane. In the Born (OPE) approximation with
electron mass neglected, $P_y=Im(T_{11})=0$. The tensor (quadrupole)
polarizations can be obtained from two ways. The first way is to use
the unpolarized electron beam and the polarized deuteron target. The
tensor polarizations are given by the ratio between the cross
section with the definite polarizations of the deuteron target and
the unpolarized cross section. Another way is to measure the
polarization of the recoiled deuteron with the unpolarized beam and
target \cite{Arnold80}. A detailed discussion about the
polarizations can be found in \cite{Ohlsen72}.
\\

The polarization which has been most widely discussed is $P_{zz}$
($T_{20}$). In the present general case, it is given as
\begin{eqnarray}
-I_0 P_{zz}&=&\frac{8}{3}\tau (G_CG_Q)+\frac89\tau^2 G_Q^2
+\frac13\tau\Big [1+2(1+\tau)\tan^2\frac{\theta}{2}\Big
]G_M^2+\Delta P_{zz},\label{I0Pzz}
\end{eqnarray}
where $\Delta P_{zz}=\delta P_{zz}+\delta_0P_{zz}$ with
\begin{eqnarray}
\delta P_{zz}&=&\frac{4\tau^2}{3}\Big [ 2(2\tau+1)G_1
-(4\tau+1)G_2+4\tau(\tau+1)G_3 \Big ]Re(G_4^{(2)*})
\nonumber \\
&&+\frac43\tau \cot^2\frac{\theta}{2}\Big
[\frac{4\tau^2+2\tau+1}{\tau+1} G_1-\frac{\tau (4\tau+1)}{\tau+1}G_2
+4\tau^2G_3\Big ]Re(G_4^{(2)*})\nonumber \\
&&+\frac{2K_0}{3M}\Big [ \Big ( \frac{4\tau^2+2\tau+1}{\tau+1}G_1
-\frac{3\tau(2\tau+1)}{2(\tau+1)}G_2+4\tau^2G_3
+2\tau^2\tan^2\frac{\theta}{2}G_2\Big ) Re(G_5^{(2)*})\nonumber \\
&&+\tau \Big ( 4(2\tau+1)G_1-(8\tau+1)G_2 +8\tau(\tau+1)G_3 \Big
)Re(G_6^{(2)*})\Big ], \label{Pzz}
\end{eqnarray}
and
\begin{eqnarray}
\delta_0P_{zz} &=&\frac{8}{3}\tau \Big
[G_CRe(G^{(2)*}_Q)+G_QRe(G^{(2)*}_C)
\Big ]\nonumber \\
&&+\frac{16}9\tau^2G_QRe(G_Q^{(2)*}) +\frac23\tau\Big
[1+2(1+\tau)\tan^2\frac{\theta}{2}\Big ]G_MRe(G_M^{(2)*}).
\end{eqnarray}
Note that $\delta_0P_{zz}$ is obtained from replacing $G_i$ by
$\G_i$ in the first three terms in Eq. (\ref{I0Pzz}) and retain only
the OPE-TPE interference terms. It should be stressed that this
polarization observable is often measured to determine the three
from factors $G_{C,Q,M}$ in the literature
~\cite{Alexa99,Garcon94}, because it relates to $G_Q^2$ and $G_QG_C$.\\

The general form for the tensor polarization $P_{xz}$ (or $T_{21})$
is
\begin{eqnarray}
I_0P_{xz}=-\tau\frac{K_0}{M_D}\tan\frac{\theta}{2} G_MG_Q +\Delta
P_{xz},
\end{eqnarray}
where $\Delta P_{xz}=\delta P_{xz}+\delta_0 P_{xz}$ with
\begin{eqnarray}
\delta P_{xz}&=&\tau\bigg \{-\frac{K_0}{M_D}\Big
[\tan\frac{\theta}{2} \tau
G_2-\frac{\cot\frac{\theta}{2}}{\tau+1}\Big ( 2\tau G_1
-(3\tau+1)G_2+2\tau(\tau+1) G_3\Big ) \Big ]Re(G_4^{(2)*})\nonumber \\
&&+\tan\frac{\theta}{2}\Big [\Big ( 2\tau G_1+(2\tau-1)G_2
+2\tau(\tau+1)G_3\Big )Re(G_5^{(2)*})
-4\tau(\tau+1)G_2Re(G_6^{(2)*}) \Big ]\nonumber \\
&&+2\tau \cot\frac{\theta}{2}\Big [2\Big (
\frac{1}{\tau+1}G_1+G_3\Big  )
Re(G_5^{(2)*})\nonumber \\
&&+\Big ( G_1-4G_2+2(\tau+1)G_3\Big ) Re(G_6^{(2)*})\Big ]\bigg \},
\end{eqnarray}
and
\begin{eqnarray}
\delta_0P_{xz}=-\tau\frac{K_0}{M_D}\tan\frac{\theta}{2} \Big
[G_MRe(G_Q^{(2)*})+G_QRe(G_M^{(2)*})\Big ].
\end{eqnarray}
For the tensor polarization $(P_{xx}-P_{yy})$ (or $T_{22})$, we
obtain
\begin{eqnarray}
I_0 (P_{xx}-P_{yy})=-\tau G_M^2 +\Delta (P_{xx}-P_{yy}),
\end{eqnarray}
where $\Delta (P_{xx}-P_{yy})=\delta (P_{xx}-P_{yy})+\delta_0
(P_{xx}-P_{yy})$ with
\begin{eqnarray}
\delta (P_{xx}-P_{yy})&=&4\tau\Big [\tau G_2
+\frac{\cot^2\frac{\theta}{2}}{\tau+1}\left (G_1+\tau G_2\right
)\Big ]
Re(G_4^{(2)*})\nonumber \\
&&+\frac{2K_0}{M_D}\Big[\Big ( \frac{1}{\tau+1}G_1
+\frac{\tau}{\tau+1}G_2\Big ) Re(G_5^{(2)*})+\tau
G_2Re(G_6^{(2)*})\Big],
\end{eqnarray}
and
\begin{eqnarray}
\delta_0(P_{xx}-P_{yy})=-2\tau G_MRe(G_M^{(2)*}).
\end{eqnarray}
\par\noindent\par

Since both the OPE and the TPE are included in the present
calculation, we obtain   non-vanishing polarization $P_y$. It is
given by
\begin{eqnarray}
I_0P_{y}&=&\frac23\tan\frac{\theta}{2} \bigg \{\frac{K_0}{M_D}\Big
[-(\tau+1) \Big (G_1Im(G_2^{(2)*})+G_2Im(G_1^{(2)*})\Big )
\nonumber \\
&&-\tau(\tau+1)\Big ( G_2Im(G_3^{(2)*})+G_3Im(G_2^{(2)*})\Big )
\nonumber \\
&&+\tau\Big ( \cot^2\frac{\theta}{2}(2G_1-G_2+2\tau G_3)+\tau
G_2\Big ) Im(G_4^{(2)*})\Big ]
\nonumber \\
&&+ \Big [ 2\tau \cot^2\frac{\theta}{2}(2G_1-G_2+2\tau G_3)
+\tau\Big (2(\tau+1)G_1+G_2+2\tau(\tau+1)G_3 \Big)\Big ]
Im(G_5^{(2)*})
\nonumber \\
&&+4\tau(\tau+1)\Big [ \cot^2\frac{\theta}{2}(G_1+\tau G_3)+\tau G_2
\Big ] Im(G_6^{(2)*})\bigg \}.
\end{eqnarray}

The other two vector polarizations $P_{x}$ (or $T_{11}$) $P_z$ (or
$T_{10}$) do not vanish in the OPE approximation and have not been
measured in the polarized electron deuteron scattering because they
are too small to be practical. In the presence of TPE contribution,
it can be written as
\begin{eqnarray}
I_0P_{z}=\frac13\frac{K_0}{M_D}\sqrt{\tau(\tau+1)}\tan^2\frac{\theta}{2}
G_M^2+\Delta P_{z},
\end{eqnarray}
where $\Delta P_z=\delta P_z+\delta_0 P_z$ with
\begin{eqnarray}
\delta P_{z}&=&-\frac{2\tau}{3}\sqrt{\frac{\tau}{\tau+1}}
\Big [\frac{K_0}{M_D}G_2Re(G_4^{(2)*}) \nonumber \\
&&+\Big ( 3+2(\tau+1)\tan^2\frac{\theta}{2}\Big ) G_2Re(G_5^{(2)*})
+2(\tau+1)G_2Re(G_6^{(2)*})\Big ],
\end{eqnarray}
and
\begin{eqnarray}
\delta_0 P_{z}&=&\frac23
\frac{K_0}{M_D}\sqrt{\tau(\tau+1)}\tan^2\frac{\theta}{2}G_MRe(G_M^{(2)*}),
\end{eqnarray}
and we obtain the single polarization $P_{x}$
\begin{eqnarray}
I_0P_{x}&=& -\frac{4}{3}\sqrt{\tau(\tau+1)}\tan\frac{\theta}{2}G_M
\Big (G_c+\frac13\tau G_Q\Big ) +\Delta P_{x},
\end{eqnarray}
where $\Delta P_x=\delta P_x+\delta_0 P_x$ with
\begin{eqnarray}
\delta P_{x}&=&\frac13\sqrt{\tau(\tau+1)} \tan\frac{\theta}{2}\bigg
\{-4\tau^2\Big ( 1+\frac{1}{\tau+1}\cot^2 \frac{\theta}{2}\Big )
G_2Re(G_4^{(2)*})\nonumber \\
&&+\frac{K_0}{M_D}\Big [\Big ( 2G_1-\frac{4\tau+1}{\tau+1}G_2 +2\tau
G_3\Big )Re(G_5^{(2)*})-4\tau G_2Re(G_6^{(2)*})\Big ]\bigg \},
\end{eqnarray}
and
\begin{eqnarray}
\delta_0 P_{x}&=&-\frac43\sqrt{\tau(\tau+1)}
\tan\frac{\theta}{2}\Big [G_M\Big (Re(G_c^{(2)*})+\frac{1}{3}\tau
Re(G_Q^{(2)*})\Big )
\nonumber \\
&+&(G_c+\frac{1}{3}\tau G_Q)Re(G_M^{(2)*})\Big ].
\end{eqnarray}
\par\noindent\par

\section{Discussions}
\par\noindent\par

The extractions of the deuteron form factors have so far been
carried out within the OPE approximation. The motivation of
investigating the $eD$ scattering beyond the OPE approximation is to
identify the possible contribution of the TPE. The extractions of
the deuteron form factors would have to be modified when the
multiple-photon exchange effects become non-negligible. As a result
it is important to know how and where the TPE effects will
begin to play a role  in the $eD$ scattering.\\

A simple and straightforward way to identify the TPE signature is to
find some constraints between observables in the framework of the
OPE approximation. Any deviation from these constraints would arise
from the TPE effects. In particular, if we can find some
$\theta$-independent combinations of physical observables then it
will not be difficult to check them experimentally. We have found
two such quantities, namely,
\begin{eqnarray}
{\cal C}_{1}&=&I_{0}(1+2P_{zz})=G_{C}^{2}-\frac{16}{3}\tau
G_{C}G_{Q}
-\frac{8}{9}\tau^2G_{Q}^{2}, \label{C1}  \\
{\cal C}_{2}&=&\frac{(I_{0}P_{xz})(I_{0}P_{x})}{I_{0}P_{z}} =4\tau
G_{Q} \left(G_{C}+\frac{\tau}{3}G_{Q}\right ). \label{C2}
\end{eqnarray}
In Eqs. (\ref{C1}-\ref{C2}),  the {\it r.h.s.} depend on $Q^2$ only.
However, $I_{0}, P_{zz}, P_{xz}, P_{x}$, and $P_{z}$ on the {\it
l.h.s.} depend both on $\theta$ and $Q^2$ but their
$\theta$-dependence cancels out in these two particular
combinations. Note that in Eqs. (\ref{C1}-\ref{C2}), we do not
express $I_{0}$ in terms of $A$ and $B$ because such a separation
holds only within the OPE framework. All the quantities which enter
in Eqs. (\ref{C1}-\ref{C2}) are directly measurable experimentally
with no need of the separation between OPE and TPE contributions.
Any $\theta$-dependence exhibited in ${\cal C}_{1}$ and ${\cal
C}_{2}$ at any fixed value of $Q^2$ would be a clear indication of
the two-photon exchange effects. In particular, should  be possible
to check Eq.~(\ref{C1}) with the existing data. It will also serve
as a good check for any theoretical prediction. In addition, it is
easy to see that $ {\cal C}_{1}+\frac{4}{3}{\cal
C}_{2}=G_{c}^{2}+\frac{8}{9}\tau^2 G_{Q}^{2}>0.
$\\

Furthermore, it is also possible to derive some constraints at some
specific angles. For example, when $\theta=\pi/2$ one can easily
derive three constraints as the  following,
\begin{eqnarray}
&&P_{z}+\frac{2}{3}\sqrt{(\tau+1)(\tau+2)}(P_{xx}-P_{yy})=0,\label{d1} \\
&&(2\tau+3)P_{z}^{2}-\frac{\tau}{6}\sqrt{(\tau+1)(\tau+2)}
P_{xz}P_{x}\nonumber\\
&&\hspace{3.0cm}+\frac{\tau^2}{6}(\tau+1)P_{xz}^{2}
+\frac{3}{8}(\tau+2)P_{x}^{2}-\sqrt{(\tau+1)(\tau+2)}P_{z}=0,\label{d2} \\
&&(2\tau+3)P_{z}^{2}-\frac{4\tau}{3}\sqrt{(\tau+1)(\tau+2)}P_{xz}P_{x}
-2\sqrt{(\tau+1)(\tau+2)}P_{z}P_{zz}=0. \label{d3}
\end{eqnarray}
Here all observables are measured at $\theta=\pi/2$. Similar
constraints at other angles such as $\theta= 2\pi/3$ or
$\theta=\pi/3$ can also be found but with more complicated structure
since $\tan\, (\theta/2)=\cot\, (\theta/2)=1$ when
$\theta=\pi/2$.\\

In the previous section, we have presented the most general forms
for the differential cross section and polarizations in $eD$ elastic
scattering. They are expressed in terms of six form factors which
are the complex functions of $Q^2$ and $\epsilon$. We see that the
TPE effects provide different $\theta$-dependence from the OPE in
all observables. By analyzing the $\theta$-dependence of the
kinematical pre-factors which appear with the interference terms of
$G_i\G_j^{(2)*}$ with ($i=1-3$, and $j=1-6$), one is able to obtain
some useful information about the TPE effects if $G_{i}^{(2)}$ are
slowly varying functions of $\theta$. More specifically, these
general expressions are greatly simplified at the forward and
backward angles limitsaoc which enable us
to disentangle the TPE effects from the OPE ones.\\

We first look at the unpolarized differential cross section. The
differential cross section at a small scattering angle $\theta\sim
8^0$ has been parametrized in Ref.~\cite{Rekalo99}  as
\begin{eqnarray}
\frac{d\sigma}{d\Omega}=(\sigma_0\cot^2\frac{\theta}{2})\bigg
|_{\theta=8^0} \frac{a_1}{(1+q^2/a_2)^{a_3}}\label{param8}
\end{eqnarray}
with parameters $a_{1,2,3}$. The $\cot^2(\theta/2)$-dependence of
this parametrization derives from OPE, as  seen in Eq.
(\ref{sigma-Mott}). In the small scattering angle region, say
$\theta\leq 15^0$, $\cot^2(\theta/2)\geq 55>>1$, We have
$\tan^2(\theta/2)\sim 1.9\times 10^{-2}\sim 0$, and $K_0\sim
2M_D\sqrt{\tau}\cot(\theta/2)$. From Eq. (\ref{Delta-sigma}), if
terms with $\tan^2(\theta/2)$-dependence are neglected, we may then
write,
%\newpage
\begin{eqnarray}
\Delta\sigma&\approx &\frac{a_1}{(1+q^2/a_2)^{a_3}}-\Big
[A(Q^2)+\Delta A(\theta,Q^2)\Big ]
\nonumber \\
&=& \frac{1}{3} \bigg \{4\tau \cot^2\frac{\theta}{2}\Big [
(2\tau-1)G_1-2\tau G_2 +2\tau^2G_3\Big  ] Re(G_4^{(2)*})
\nonumber \\
&&+4\sqrt{\tau}\cot\frac{\theta}{2}\Big [ \left ( (2\tau-1)G_1-2\tau
G_2+2\tau^2G_3\right ) Re(G_5^{(2)*})
\nonumber \\
&&+2\tau\Big  ( (2\tau+1)G_1-(2\tau+1)G_2+2\tau(\tau+1)G_3\Big )
Re(G_6^{(2)*})\Big ]\bigg \}.\label{TPE8}
\end{eqnarray}
One thus finds that the TPE effects provide a much more complicated
$\theta$-dependence for the differential cross section compared to
Eqs.~(\ref{diffCrx},\ref{I0-ope},\ref{sigma-Mott}) as obtained
within the OPE approximation . In addition, that $\Delta\sigma$
should remain finite when $\theta$ approaches zero would require
\begin{eqnarray}
%Re\, G_{1}^{(2)} (\theta, Q^2)&\leq& \theta^0,\,\,\,Re\,
%G_{2}^{(2)}(\theta,Q^2) \leq \theta^{0},\,\,\,
%Re\, G_{6}^{(3)} (\theta,Q^2)\leq\theta^{0},\nonumber \\
Re\, G_{4}^{(2)} (\theta, Q^2)&\leq& \theta^2,\,\,\,Re\,
G_{5}^{(2)}(\theta,Q^2) \leq \theta,\,\,\,\,\,\, Re\, G_{6}^{(2)}
(\theta,Q^2)\leq\theta,\label{eq:cond}
\end{eqnarray} when $\theta\rightarrow 0$.\\

Arguments similar to those presented above can also be applied to
the polarization observables. At small angles, we obtain the
following simplified results for the tensor polarizations
\begin{eqnarray}
\Delta P_{zz}&\sim& \frac43\tau \cot^2\frac{\theta}{2}\Big [
\frac{4\tau^2+2\tau+1}{\tau+1} G_1-\frac{\tau(4\tau+1)}{\tau+1}G_2
+4\tau^2G_3\Big  ] Re(G_4^{(2)*})
\nonumber \\
&&+\frac43\sqrt{\tau}\cot\frac{\theta}{2}\Big [ \Big (
\frac{4\tau^2+2\tau+1}{\tau+1}G_1
-\frac{3\tau(2\tau+1)}{2(\tau+1)}G_2+4\tau^2G_3\Big ) Re(G_5^{(2)*})
\nonumber \\
&&+\tau\Big ( 4(2\tau+1)G_1-(8\tau+1)G_2+8\tau(\tau+1)G_3\Big )
Re(G_6^{(2)*})\Big ],  \\
\Delta P_{xz}&\sim& \tau\bigg \{ \Big  [
2\sqrt{\tau}\cot^2\frac{\theta}{2}\Big (\frac{2\tau}{\tau+1}G_1
-\frac{3\tau+1}{\tau+1}G_2+2\tau G_3\Big )\Big ] Re(G_4^{(2)*})\nonumber \\
&&+2\tau \cot\frac{\theta}{2}\Big [\Big (
\frac{1}{\tau+1}G_1+2G_3\Big ) Re(G_5^{(2)*})
\nonumber \\
&&+\Big ( G_1-4G_2+2(\tau+1)G_3\Big ) Re(G_6^{(2)*})\Big ]\bigg
\},\\
\Delta (P_{xx}-P_{yy}) &\sim& \frac{4\tau}{\tau+1}
\cot^2\frac{\theta}{2}\left ( G_1+\tau G_2\right ) Re(G_4^{(2)*})
\nonumber \\
&&+\frac{4\sqrt{\tau}}{\tau+1}\cot\frac{\theta}{2}\Big [\left ( G_1
+\tau G_2\right ) Re(G_5^{(2)*})+\tau(\tau+1) G_2Re(G_6^{(2)*})\Big
].
\end{eqnarray}
We find that the TPE effects in $P_{zz}, P_{xz}$ and $P_{xx}-P_{yy}$
are similar to the ones in the differential cross section.
\\

For vector polarizations  $P_{x}$ and $P_{z}$, we obtain at the
small angle limit
\begin{eqnarray}
\Delta P_{z}&\sim&-\frac{2\tau}{3}\sqrt{\frac{\tau}{\tau+1}} \Big
[2\sqrt{\tau}\cot\frac{\theta}{2}Re(G_4^{(2)*})+3Re(G_5^{(2)*})
+2(\tau+1)Re(G_6^{(2)*})\Big ]G_2. \nonumber \\
\Delta P_{x}&\sim&\frac{2\tau\sqrt{\tau+1}}{3} \Big [\Big (
2G_1-\frac{4\tau+1}{\tau+1}G_2+2\tau G_3\Big ) Re(G_5^{(2)*}) -4\tau
G_2Re(G_6^{(2)*})\Big ]\nonumber \\
&&-\frac43\frac{\tau^2\sqrt{\tau}}
{\sqrt{\tau+1}}\cot\frac{\theta}{2}G_2Re(G_4^{(2)*}).\label{DPx}
\end{eqnarray}
It is interesting to see the difference between $P_{x}, P_{z}$ and
other observables. In the extreme forward limit, both $P_{x}$ and
$P_{z}$ vanish  while the other observables would remain finite.
From Eqs. (\ref{eq:cond}) and (\ref{DPx}), one also sees that the
TPE effects in $P_{x}$ and $P_{z}$ decrease faster than the
corresponding effects in other observables which makes $P_{x},
P_{z}$
less interesting as far as the TPE effects are concerned.\\

%XXXXXXXXXXXXXXXXXXXXXXXXXXXXXXXXXXXXXXXXXXXXXXXXXXXXXXXXXXXXXXX

In the large angle limit when the angle approaches $\pi$, i.e., very
backward direction, expressions for the observables in $eD$ elastic
scattering are also simplified if $G_{i}^{(2)} $'s are slowly
varying functions of $\theta$. However the Mott cross section is
suppressed by $(\delta\theta)^2=(\pi-\theta)^2$, therefore no
constraints such as Eq. (\ref{eq:cond}) can be deduced. As long as
$Re\,G_{i}^{(2)}$'s $(i=1-6)$ are analytical functions of
$\delta\theta$, the observables are all finite when $\theta$
approach $\pi$. As a matter of fact, only the differential cross
section, $P_{zz}$ and $P_{z}$ will receive nonvanishing TPE
contribution in the backward directions. For example, $\theta\geq
165^0$, then $\cot^2(\theta/2)\leq 1.9\times 10^{-2}\sim 0$ and
$\tan^2(\theta/2)> 55>>1$, such $x\sim 1$ and $K_0\simeq
2M\sqrt{\tau(1+\tau)}$.  For the unpolarized differential cross
section in this limit, we then have
\begin{eqnarray}
\frac{d\sigma}{d\Omega}\bigg |_{\theta\rightarrow \pi}
\sim\sigma_0\big( B+\Delta B'\big ),
\end{eqnarray}
with
\begin{eqnarray}
\Delta B'=-\frac83\tau\sqrt{\tau(1+\tau)}G_2(Q^2)Re(G_5^{(2)*})
+\frac83\tau(1+\tau)G_2Re(G_2^{(2)*})\label{dbpi}.
\end{eqnarray}
For $P_{zz}$ (or $T_{20}$) and $P_z$ (or $T_{10}$) one obtains
\begin{eqnarray}
\Delta P_{zz}\sim
\frac83\tau^2\sqrt{\tau(\tau+1)}\tan^2\frac{\theta}{2}G_2Re(G_5^{(2)*})
+\frac43\tau(\tau+1)\tan^2\frac{\theta}{2}G_2Re(G_2^{(2)*}),
\label{dpzzpi}
\end{eqnarray}
and
\begin{eqnarray}
\Delta
P_{z}&\sim&\frac43\sqrt{\tau(\tau+1)}\tan^2\frac{\theta}{2}\Big
[\sqrt{\tau(\tau+1)}Re(G_M^{(2)*})-Re(G_5^{(2)*})\Big ]G_2.
\label{dpzpi}\end{eqnarray} We note that  the form factors which
appear in Eqs. (\ref{dbpi}-\ref{dpzpi}) are $Re(G_{2}^{(2)})$ and
$Re(G_{5}^{(2)})$.\\

 The vector polarization $P_{y}$ is unique
in that it is related to the imaginary parts of the form factors. In
the small angle limit, we may write
\begin{eqnarray}
I_0P_{y}&\sim&\frac23\bigg \{2\tau\sqrt{\tau}\cot^2\frac{\theta}{2}
\Big [2G_1-G_2+2\tau G_3\Big ]Im(G_4^{(2)*})
\nonumber \\
&&+2\tau \cot\frac{\theta}{2}\Big [(2G_1-G_2+2\tau
G_3)Im(G_5^{(2)*})
\nonumber \\
&&+2(\tau+1)(G_1+\tau G_3)Im(G_6^{(2)*})\Big ]\bigg \}.
\end{eqnarray}
If $I_{0}P_{y}$ remain finite at small angles then the following
conditions have to be satisfied:
\begin{eqnarray}
Im\,G_{4}^{(2)}(\theta, Q^2)\leq \theta^2,\,\,\,
Im\,G_{5}^{(2)}(\theta,Q^2)\leq \theta,\,\,\,
Im\,G_{6}^{(2)}(\theta,Q^2)\leq\theta,\label{eq:cond2}\end{eqnarray}
  when
$\theta\rightarrow 0$. If Eq. (\ref{eq:cond2}) is satisfied then
$P_{y}$ receives nonzero TPE effects at small angles. Since $P_y$
vanishes in OPE approximation, any measurement which yields a
nonzero value of $P_y$ would be a manifestation
of the TPE effects.\\

\section{Summary}

\par\noindent\par

In summary, we considered the elastic $eD$ scattering beyond the
one-photon-exchange approximation. The scattering amplitude is
expressed in terms of six Lorentz structures as dictated by the
charge-conjugation and parity invariance, together with  helicity
conservation. The six invariant functions $G_{i}$'s, $i=1-6$ are
complex functions of $Q^2$ and, e.g., $\theta$. We derived the
general expressions for the complete set of seven observables,
including the differential cross section, three vector polarizations
and three tensor polarizations, in terms of the bilinear products
$G_{i}G^{*}_{j}, \,\,i=1-3, \,\,j=1-6$, where $G_j=0$ for $j=4-6$ in
the one-photon approximation. The general expressions which include
the two-photon exchange effects were examined in details. \\

We found two $\theta$-independent relations as given in Eqs.
(\ref{C1}-\ref{C2}), for any fixed value of $Q^2$ between these
observables within the OPE approximation. Any deviation from these
relations found in the experiments would be a clear indication of
the two-photon-exchange mechanism. This is very useful since it will
provide us with the much needed information about the precise
kinematical region, i.e., values of momentum transfer square $Q^2$
and the scattering angle $\theta$, where the TPE effects would show
up. In addition,  we derived three relations in Eqs.
(\ref{d1}-\ref{d3}), between polarization observables at
$\theta=\pi/2$. They also serve as useful
 criteria on the
validity of the OPE approximation in the $eD$ scattering.\\

Finally, we discussed the possibilities of observing the TPE effects
in some special kinematical regions, such as the forward and
backward angles limit, under the assumption that the form factors
$G_i$'s are slowly  varying functions of $\theta$. Whether such an
assumption holds has to be substantiated by theoretical
consideration. A model calculation in this direction is currently
underway \cite{zhou06}.
\\

\section*{Acknowledgements}
\par\noindent\par\noindent\par
The work of Y.B.D. is supported in part by the National Sciences
Foundations of China under grant No. 10475088. He would also like to
thank Chinese Development Fund of Taiwan for support during his
visit to the Department of Physics, National Taiwan University where
he was warmly received. The work of C.W.K. is supported in part by
the National Science Council of Taiwan  under grant No.
NSC95-2112-M033-014. The work of S.N.Y. and Y.C.C. is supported in
part by the National Science Council of Taiwan  under grant No.
NSC95-2112-M002-025. We like to thank Haiqing Zhou for useful
discussions.

\end{document}